# Radio Frequency Magnet-free Circulators Based on Spatiotemporal Modulation of Surface Acoustic Wave Filters

Yao Yu, *Student Member, IEEE,* Giuseppe Michetti, Michele Pirro, Ahmed Kord, *Student Member, IEEE,* Dimitrios L. Sounas, *Senior Member, IEEE*, Zhicheng Xiao, *Student Member, IEEE*, Cristian Cassella, *Member, IEEE,* Andrea Alù, *Fellow, IEEE*, and Matteo Rinaldi, *Senior Member, IEEE*.

*Abstract*—In this paper, a new generation of magnet-free circulators with high performance is proposed. Circulators are crucial devices in modern communication systems due to their ability to enable full-duplexing and double the spectral efficiency directly in the physical layer of the radio-frequency (RF) front-end. Traditionally, Lorentz reciprocity is broken by applying magnetic bias to ferrite materials, therefore conventional circulators are bulky and expensive. In this paper, this problem is addressed by replacing the magnetic bias with periodic spatiotemporal modulation. Compared to previous works, the proposed circulator is constructed using surface acoustic wave (SAW) filters instead of transmission lines (TL), which reduces the modulation frequency by at least a factor of 20 and ensures ultra-low power consumption and high linearity. The miniaturized high quality (Q) factor SAW filters also lead to a low-loss non-reciprocal band with strong isolation (IX) and broad bandwidth (BW) on a chip scale, therefore addressing such limitations in previous magnet-free demonstrations. Furthermore, compared to the conventional differential circuit configuration, a novel quad configuration is developed, which doubles the intermodulation-free bandwidth.

## I. INTRODUCTION

In modern communications, full-duplex operation can in principle double the spectrum efficiency of traditional half-duplex systems [1]-[7]. In such systems, circulators are crucial to transmit the RF signal in one direction while blocking it in the opposite direction. Traditionally, these non-reciprocal devices require a strong magnetic bias in order to break time-reversal symmetry, leading to bulky and expensive device, which partially contribute to the long-held false assumption that full-duplex systems are impractical. Transistors have been considered as potential candidates to build magnet-free circulators [8]-[10], however, they lead to fundamental limitations in terms of linearity and noise performance. Recently, a new class of magnet-free circulators based on time-varying circuits was proposed [11]-[35], reviving the hope to realize fully integrated full-duplex systems in the near future. In time-varying circulators, the magnetic bias is replaced by periodic spatiotemporal modulation, thus having the potential to achieve non-reciprocity at a much smaller form factor. Varactors are considered as suitable components to apply periodic modulation [11], [12]. Nevertheless, the use of varactors comes with some fundamental challenges, such as poor linearity and complicated modulation networks, therefore varactors-based circulators are limited in real applications which require an integrated circulator system with high power handling. On the other hand, RF switches can provide a better linearity and alleviate the necessity of using complicated modulation networks, therefore they can overcome the limitations that varactors suffer from. However, a trade-off between switching speed and linearity imposes a stringent requirement on the modulation frequency. In [15] and [16], TLs have been periodically modulated to break reciprocity, however, due to the limited loaded Q factor, large modulation frequencies are required, thus small RF switches with fast switching speed are used, resulting in challenges in terms of linearity. Even though [15] improves the linearity by suppressing the voltage swing at receiver port, this results in an asymmetrical circulator response, leaving the network sensitive to impedance mismatches. In fact, all the aforementioned demonstrations of magnet-free circulators, either based on varactors or RF switches, face the challenge of requiring large modulation frequencies or narrow BW of operation. Besides poor linearity and leakage of RF power into the modulation network, the large modulation frequency also leads to large power consumption, which is another critical metric for communication systems.

In this paper, we consider SAW filters periodically

Manuscript received May 28, 2019. This work was supported by the Defense Advanced Research Projects Agency (DARPA) SPAR program under contract no. HR0011-17-2-0002.

Y. Yu, G. Michetti, M. Pirro, C. Cassella and M. Rinaldi are with the SMART center, Northeastern University, Boston, MA 02115 USA. (corresponding author: M Rinaldi. Phone: 617-373-2751; Fax: 617-373-8970; E-mail: rinaldi@ece.neu.edu).

A. Kord and Z. Xiao are with the Department of Electrical and Computer Engineering, University of Texas at Austin, Austin, TX78712, USA.
D. L. Sounas is with the Department of Electrical and Computer Engineering, Wayne State University, Detroit, MI 48202, USA.
A. Alu is with the Department of Electrical and Computer Engineering, University of Texas at Austin, TX78712, USA, and also with the Advanced Science Research Center, City University of New York, New York, NY 10031, USA.

modulated in time to break reciprocity. Due to the energy conversion between electrical and mechanical domains, high Q factors are guaranteed, thus resulting in at least 20 times smaller modulation frequency and more than 850 times reduction in power consumption compared to other implementations working at similar center frequency [24]. The low modulation frequency enables the use of large RF switches with high linearity. As a result, one of the highest linearity among all the magnet-free circulators is achieved. The high Q factor and constant group delay of the SAW filters also guarantee a strong non-reciprocal band with low insertion loss (IL) and broad BW. Furthermore, in order to improve the usable BW without intermodulation products (IMPs), a new circuit configuration, named quad configuration, is proposed. Compared to the conventional differential configuration, at which the IMPs closest to the fundamental tones are the second-order ones, this quad configuration cancels the second-order IMPs, thus doubling the usable bandwidth compared to all previous demonstrations of time-varying circulators. As a result, we present a time-varying circulator with ultra-low modulation frequency and power consumption, showing at the same time low IL, strong IX, broad BW and high linearity, thus addressing the major challenges that currently prevent the realization of magnet-free integrated full-duplex components with high performance.

## II. DESIGN

The proposed circuit architecture is shown in Fig. 1(a). Two SAW filters are periodically modulated through RF switches, controlled using square waves $M_1$ and $M_2$ with a 50% duty cycle and a phase of $\varphi = 0^o$ and $90^o$, respectively. The modulation period $T_m$ of the control signals are set to be four times the group delay of the SAW filters, i.e., $T_m = 4T_g$, where $T_g$ is the group delay, so that the time delay between $M_1$ and $M_2$ is equal to the group delay. Each port is connected to a pair of complementary switches $SW_{i1}$ and $SW_{i2}$, where $i$ is the port number, forming a differential configuration. The operation of the circuit can be explained through the time diagram shown in Fig. 1(b)-(c). When port 1 is excited as shown in Fig. 1(b), $SW_{11}$ ($\varphi = 0^o$) is turned on from $t = 0$ to $t = \frac{T_m}{2}$, allowing the incident signal to be transmitted to the upper SAW filter during this time window. As the signal passes through the SAW filter, it exhibits a delay of $T_g$, then it reaches $SW_{21}$ ($\varphi = 90^o$) which also has a time delay of $\frac{T_m}{4} = T_g$ with respect to $SW_{11}$. Therefore, the signal is transmitted through $SW_{21}$ as well, and it is delivered to port 2. Similarly, from $t = \frac{T_m}{2}$ to $t = T_m$, the signal is transmitted to port 2 through the lower branch. Assuming that the filters have infinite BW and dispersion-free group delay, all the power excited from port 1 is delivered to port 2. On the other hand, if the signal is excited from port 2, it can be shown that all the incident power is transmitted to port 3, as depicted in Fig. 1(c). The same time diagram can also be drawn for the signal transmission from port 3 to port 4. Therefore, the input signal will circulate in the following order 1→2→3→4→1, which is the operation of a circulator.

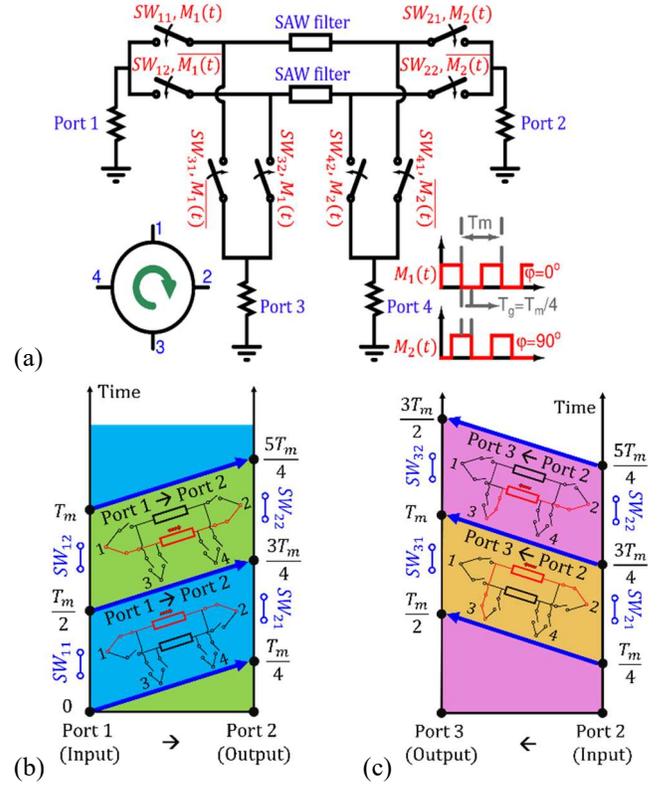

Fig. 1 (a) Circuit schematic of the proposed circulator. Left inset: signal transmission direction of the circulator schematic. Right inset: control signals in time domain. (b) Input signal is excited from port 1 and transmitted to port 2. All the other ports have no power transmission. (c) Input signal is excited from port 2. Instead of going back to port 1, the signal is transmitted to port 3, showing non-reciprocity.

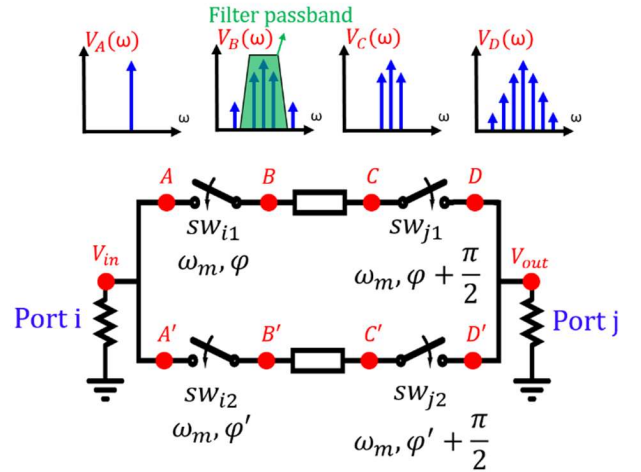

Fig. 2. Schematic of the input and output ports. Voltage spectra at different points from the input port to the output port are labeled. The two isolated ports are hidden for simplicity.

The time diagram in Fig. 1 is based on the assumption that the SAW filters have infinite BW, which is physically unrealistic. After the input signal is mixed with the clock of the switch $SW_{i1}$ or $SW_{i2}$ (i=1, 2, 3 or 4), infinite numbers of IMPs will be generated at points B and B', depicted in Fig. 2. Due to the finite BW of the SAW filters, only a finite number of IMPs will be transmitted. For the SAW filter experimentally demonstrated in this paper (IL, RL and group delay are shown in Fig. 3), only the first-order IMPs fall inside the BW. However, as will be shown next, most of the signal power is in fact carried out by the fundamental tone and the first-order IMPs, therefore the filtering process described above will cause only a small IL to the system.

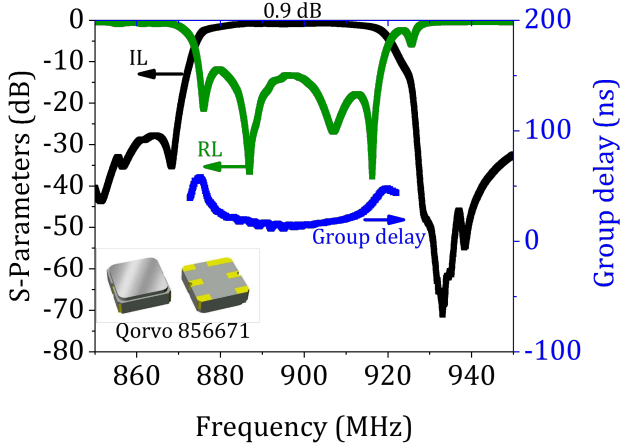

Fig. 3. The IL, RL and group delay of the SAW filter Qorvo 856671.

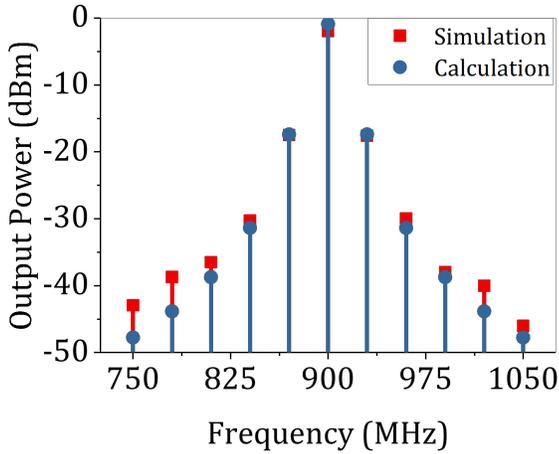

Fig. 4. Simulated and calculated output spectra of the schematic shown in Fig.2.

Fig. 2 depicts the generation of IMPs from the input to the output port, i.e., i = 1, 2, 3 or 4 and j = 2, 3, 4 or 1, respectively. As shown in Fig. 2, the voltage swing at the input has a frequency $\omega_{RF}$ and its spectrum can be written as

$$V_A(\omega) = \delta(\omega - \omega_{RF}) \quad (1)$$

where δ denotes the Dirac delta function. After mixing with the switching signal of $SW_{i1}$ ($M_{i1}(t)$), the voltage at point B becomes the convolution of $V_{in}$ and a square wave with 50% of duty cycle:

$$V_B(\omega, \varphi) = V_{in}(\omega) * \widehat{M}_{i1}(\omega, \varphi) \quad (2)$$

where $\widehat{M}_{i1}(\omega, \varphi)$ is the Fourier transformation of the square wave clock signal $M_{i1}(t)$ and $\varphi$ is the phase of the modulation signal. For a 50% duty cycle square wave, $\widehat{M}_{i1}(\omega, \varphi)$ can be written as follows

$$\widehat{M}_{i1}(\omega, \varphi) = \widehat{M}_{i1}(\omega, \varphi = 0) \times e^{-i\varphi(\frac{\omega}{\omega_m})} \quad (3)$$

where

$$\widehat{M}_{i1}(\omega, \varphi = 0) = \frac{1}{2}\delta(\omega) + \sum_{n=1}^{\infty}\{\frac{i}{n\pi}\left|\sin\left(\frac{n\pi}{2}\right)\right|[\delta(\omega + n \cdot \omega_m) - (\delta(\omega - n \cdot \omega_m)]\} \quad (4)$$

where $\omega_m$ is the radius modulation frequency. Therefore, point B will see an infinite numbers of IMPs. Assuming that the SAW filter has a brick-wall passband response with zero IL, constant group delay $T_d$, and a BW that can transmit only the fundamental tone and the first-order IMPs, all the higher-order IMPs will be filtered out and the voltage at point C will only have three frequency components, i.e.,

$$V_C(\omega, \varphi) = V_B(\omega, \varphi) \times \begin{cases} e^{-i\omega T_d} & \omega_{RF} - \omega_m \leq \omega \leq \omega_{RF} + \omega_m \\ 0 & elsewhere \end{cases} =$$

$$a_1 \cdot \delta(\omega - \omega_{RF} - \omega_m) + a_2 \cdot \delta(\omega - \omega_{RF}) + a_3 \cdot \delta(\omega - \omega_{RF} + \omega_m) \quad (5)$$

Substituting Eq. (2)-(4) into Eq. (5), the amplitudes of these three frequency components can be calculated as follows

$$a_1 = -\frac{e^{-i(\omega_{RF}T_d+\varphi)}}{\pi}$$
$$a_2 = \frac{1}{2}e^{-i\omega_{RF}T_d}$$
$$a_3 = -\frac{e^{-i(\omega_{RF}T_d-\varphi)}}{\pi} \quad (6)$$

These three frequency components are then mixed with the switching signal of the second switch which exhibits a phase of $\varphi + \pi/2$, and the voltage at point D can be derived,

$$V_D(\omega, \varphi) = V_C(\omega, \varphi) * \widehat{M}_{j1}\left(\omega, \varphi + \frac{\pi}{2}\right) =$$
$$V_C(\omega, \varphi) * (\widehat{M}_{i1}(\omega, \varphi) \cdot e^{-i\frac{\pi}{2}}) =$$
$$\sum_{n=0}^{\infty} b_n \delta(\omega - \omega_{RF} - n \cdot \omega_m) \cdot e^{-i\omega_{RF}T_d} \quad (7)$$

where

$$b_n = \begin{cases} \frac{1}{4} + \frac{2}{\pi^2} & n = 0 \\ -\frac{1}{\pi}e^{-in\varphi} & n = \pm 1 \\ (-1)^{\frac{n}{2}} \cdot \frac{1}{\pi^2}\left(\frac{1}{n+1} - \frac{1}{n-1}\right) \cdot e^{-in\varphi} & n \text{ is even}, n \neq 0 \\ -\frac{1}{2n\pi}\sin\left(\frac{n\pi}{2}\right)e^{-in\varphi} & n \text{ is odd}, n \neq \pm 1 \end{cases} \quad (8)$$

Similarly, it can be shown that the voltage $V_{D'}$ has the same form as $V_D$ by replacing $\varphi$ with $\varphi'$. From Eq. (8), we observe that all the IMPs have a phase term of $e^{-in\varphi}$. Therefore, by using a differential configuration with two complementary switches, i.e., $\varphi' - \varphi = 180^o$, the output spectrum will have zero odd-order IMPs, since all the odd-order IMPs will have different signs in the output from the two differential branches, i.e., $e^{-in\varphi} = -e^{-in(\varphi+\pi)}$ when $n$ is odd, and they destructively interfere with each other. Meanwhile, even-order IMPs will have the same sign from the two branches, i.e., $e^{-in\varphi} = e^{-in(\varphi+\pi)}$ when $n$ is even, and they constructively interfere with each other, showing up at the output port. Furthermore, the fundamental tone at the output can be calculated from $V_{out}(\omega = \omega_{RF}) = V_D(n = 0) + V_{D'}(n = 0) \approx 0.905\delta(\omega - \omega_{RF}) \cdot e^{-i\omega_{RF}t_d}$. Hence, the theoretical IL, assuming zero IL and constant group delay for the SAW filter, is $-20 \times log \frac{|V_{out}|}{|V_{in}|} = 0.86\ dB$. Therefore, even though higher-order IMPs are filtered out by the SAW filter after the first switch, the filtering process will only cause a IL of 0.86 dB since most of the power is carried by the fundamental tone and first IMPs.

The output spectrum of Fig. 2 is simulated using ADS harmonic balance (Fig. 4). The S-parameters of the SAW filter Qorvo 856671 (Fig. 3) are used in obtaining these results. The simulated results are also compared to the theoretical ones calculated using Eq. (8), showing good agreement. The larger loss of the simulated fundamental tone, i.e., 1.9 dB versus 0.86 dB, is due to the non-zero IL of the SAW filter of ~0.9 dB and the group delay dispersion of 15-23 ns over the entire BW. The larger simulated values of the higher-order IMPs (n>4) can be explained by the fact that the SAW filter has a finite out-of-band rejection so higher-order IMPs at point B and B' can leak into the output of the filter (C and C').

The S-parameters of the circuit in Fig. 1(a) are simulated using ADS harmonic balance. The S-parameters of the SAW filter from Fig. 3 are used in the simulation. The switches are assumed to be resistive at both on and off states ($R_{on}$ is 5 ohm and $R_{off}$ is 1M ohm), with zero rise and fall transition times. The simulated results are shown in Fig. 5, depicting a strong non-reciprocity with an IL of 2.1 dB. The simulated BW, defined as the 20 dB-IX of $|S_{31}|$, is 30 MHz. The slightly higher IL compared to the result of the output spectrum in Fig. 4 is due to the small power leakage from port 1 to port 3 and 4 resulting from the finite dispersion of the in-band group delay.

As explained above, a differential configuration with a modulation phase $\varphi$ of $0^o$ and $180^o$ for the constituent single-ended circuits will cancel the odd-order IMPs yet maintain the even-order tones. In order to further suppress the latter, a new quad configuration is proposed, as shown in Fig. 6. At each port, four switches with a modulation phase difference of $90^o$ are used. Assuming the modulation phase of the four switches are $0^o$, $90^o$, $180^o$ and $270^o$, then the phase term $e^{-in\varphi}$ when n equals 2 will be +1 for $\varphi = 0^o$ or $180^o$, and –1 for $\varphi = 90^o$ or $\varphi = 270^o$. Therefore, the second-order IMPs from the four branches will destructively interfere with each other and the output spectrum will show zero second-order IMPs. The quad architecture is also simulated using ADS harmonic balance. The same S-parameters of the SAW filter and the resistive switch model are used in the simulation. The characteristic impedance is tuned to 25 Ohm, since at each port, the number of SAW filters is doubled. The simulated S-parameters show a strong non-reciprocity with an IL of 2.1 dB (Fig. 7). The output spectrum is also simulated (Fig. 8). As expected, all the first three IMPs are suppressed, and the closest IMPs to the fundamental tone are the fourth-order ones.

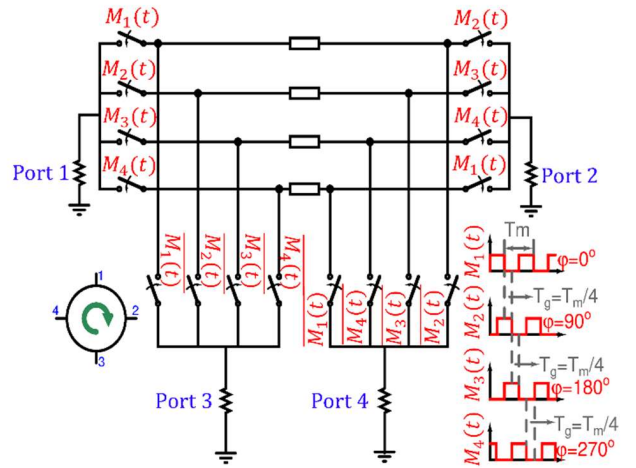

Fig. 6. Schematic of the quad configuration.

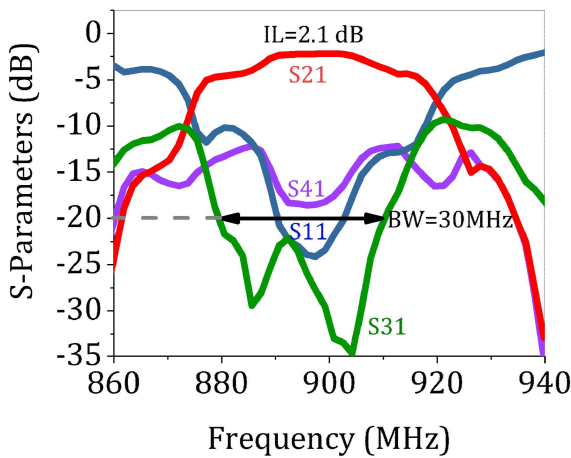

Fig. 5. Simulated S-parameters of the circulator circuit.

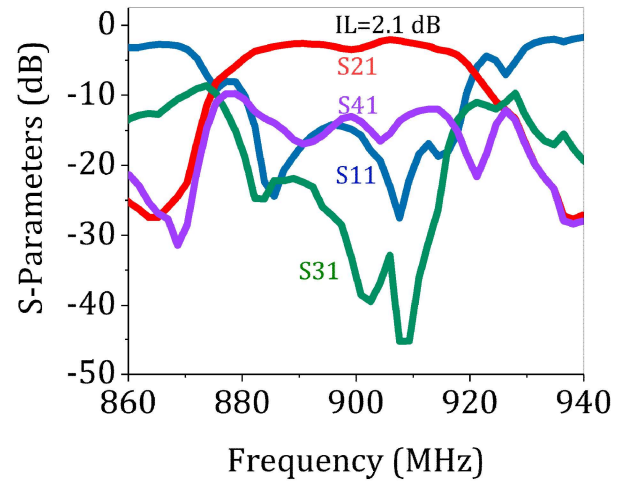

Fig. 7. Simulated S-parameters of the quad configuration.

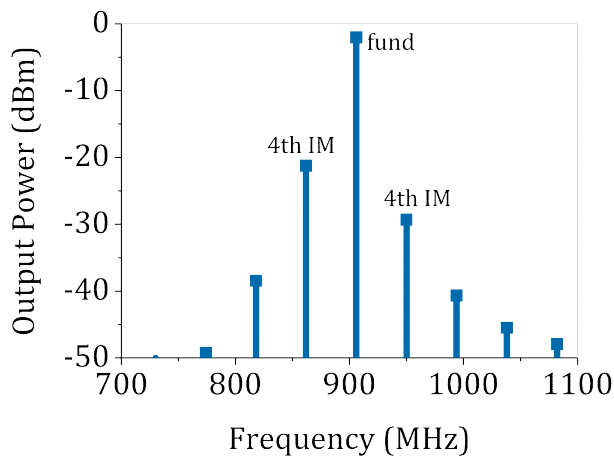

Fig. 8. Simulated output spectrum of quad configuration.

## III. Measurement

In order to test the proposed circuit architecture, a printed circuit board (PCB) prototype was designed and implemented. Fig. 9 shows a picture of the fabricated board. The SAW filters (Qorvo 856671) and the RF switches were wire-bonded to the PCB. The RF switches are engineering samples from Qorvo. Since no inductors or TLs are required in the circuit, the device area (SAW filters and switches) is only 6×6 mm$^2$.

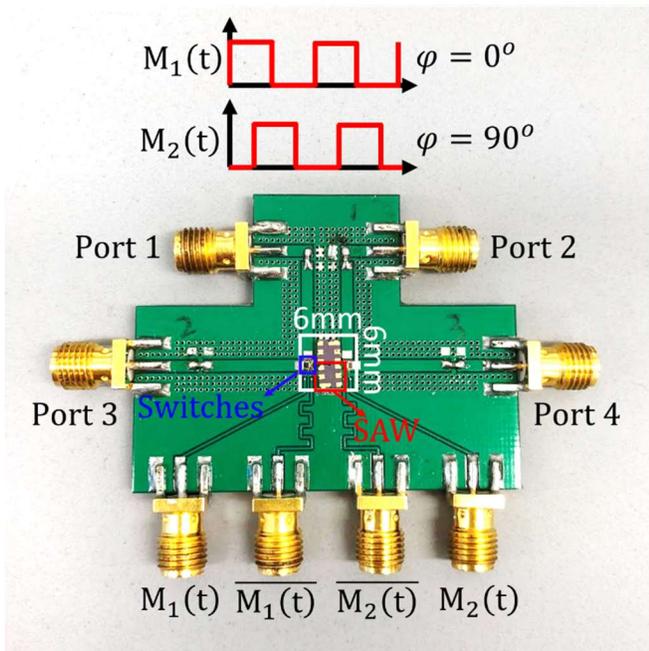

Fig. 9. PCB design and implementation.

The testing set-up is shown in Fig. 10. Two dual-channel function generators were synchronized together to provide the control signals $M_1(t)$, $\overline{M_1(t)}$, $M_2(t)$ and $\overline{M_2(t)}$. The S-parameters were measured using a four-port vector network analyzer (VNA).

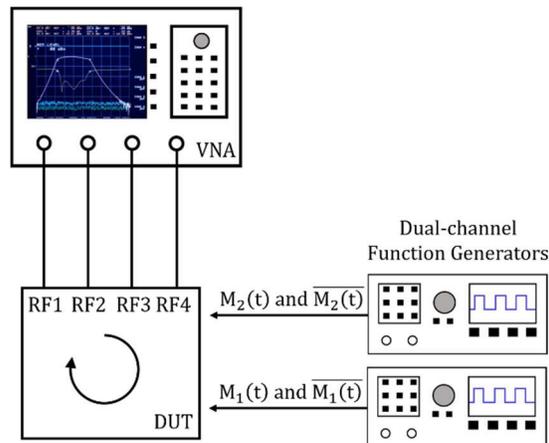

Fig. 10. The testing set up for S-parameters measurement.

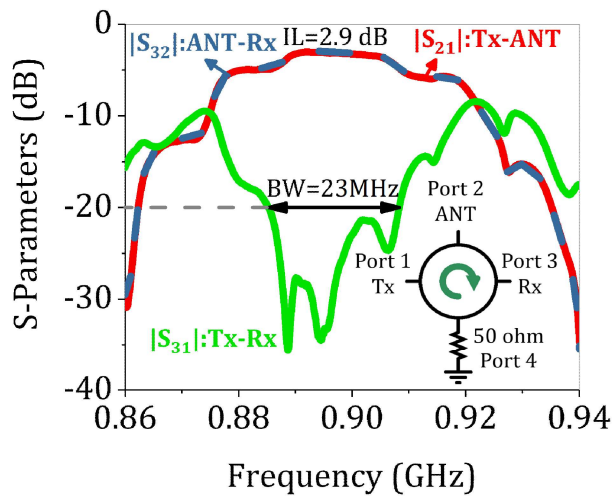

Fig. 11. Measured IL and IX of the differential configuration.

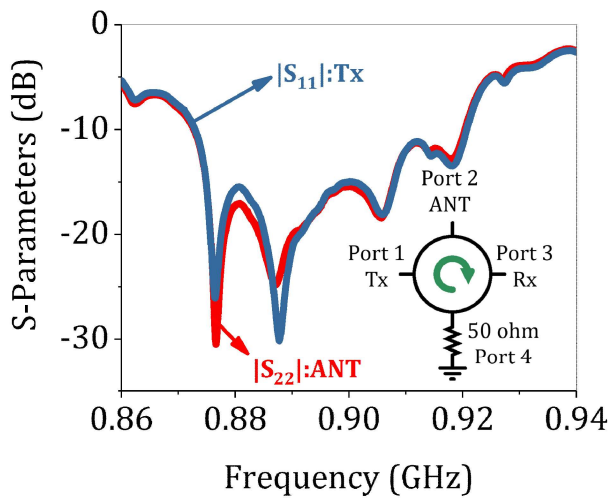

Fig. 12. Measured RL of the differential configuration.

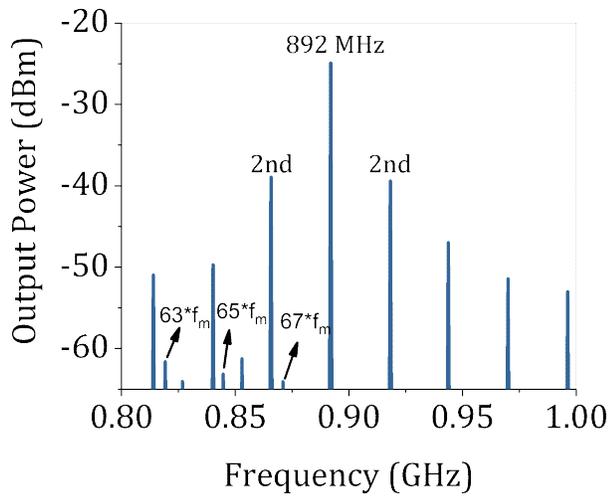

Fig. 13. Measured output spectrum.

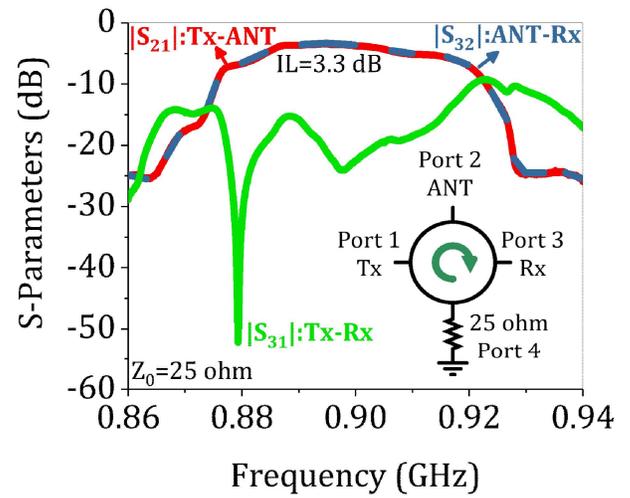

Fig. 15 Measured IL and IX for the quad configuration.

The measured S-parameters are shown in Fig. 11, 12. By assigning port 1, 2 and 3 to transmitter (Tx), antenna (ANT) and receiver (Rx), respectively, the IL of Tx-to-ANT and ANT-to-Rx, and IX of Tx-to-Rx, which are the most important metrics for the operation of full-duplex radio, are shown in Fig. 11. The measured IL is 2.9 dB. The 20 dB IX-BW is 23 MHz, i.e., ~2.6% of center frequency. The RL (Fig. 12) is larger than 15 dB over the entire BW, and can be further improved by adding a matching network or improving the RL of the filters. The slight difference between simulated and measured results is attributed to the parasitics introduced by the switches and the PCB. Specifically, the higher IL, i.e., 2.9 dB in measurement and 2.1 dB in simulation, is due to the IL from the switches of 0.23 dB each and the PCB parasitics. The smaller 20 dB IX-BW of 23 MHz in measurement and 30 MHz in simulation is attributed to the finite off-state isolation of the switches.

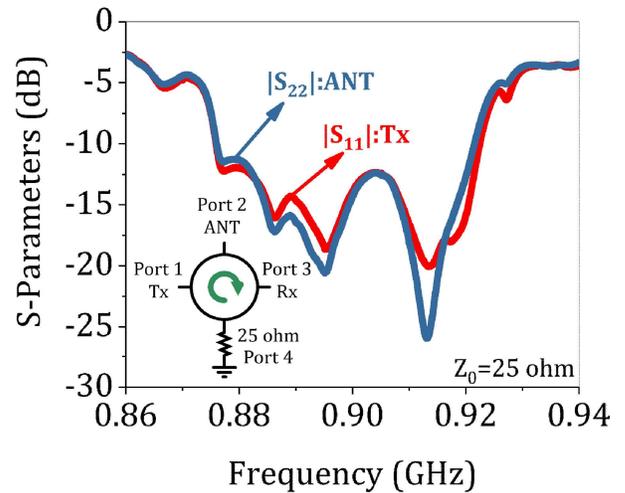

Fig. 16 Measured RL for the quad configuration.

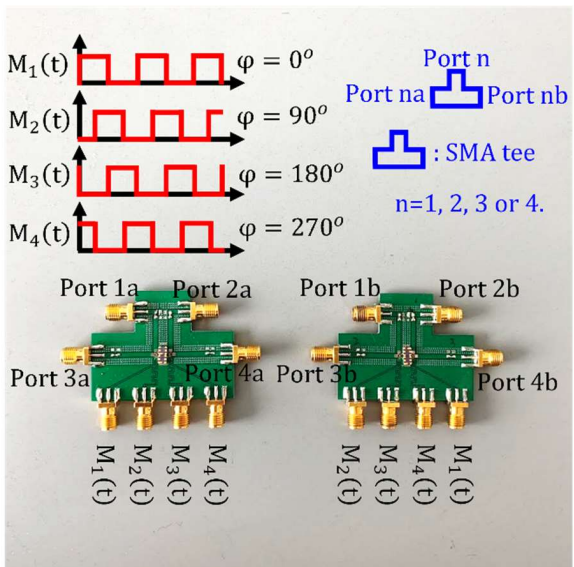

Fig. 14 The set up schematic for the implementation of the quad configuration. Two PCBs with differential configuration are connected together with proper modulation phase delay to obtain the quad configuration.

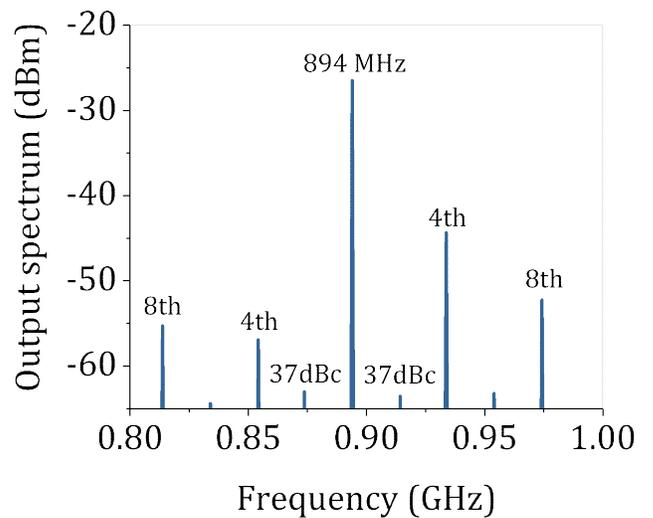

Fig. 17 Measured output spectrum for the quad configuration.

TABLE I
SUMMARIZE OF MAGNET-FREE CIRCULATORS PERFORMANCES

| | Technology | Center freq. | Mod. Freq. [a] | BW [b] | IX | IL | P1dB | IIP3 | Power Consum. |
|---|---|---|---|---|---|---|---|---|---|
| [15] | TL | 25 GHz | 33% | 18.4% | 18.3 dB | 3.3/3.2 dB | 21.5/21 dBm | N/A | N/A |
| [26] | TL | DC-3GHz [c] | 83% [d] | 93.3% [e] | 20 dB | 4.3 dB | N/A | N/A | N/A |
| [25] | LC | 950 MHz | 33% | 17% | 25 dB | 2.1/2.9 dB | 21/31 dBm | 37/50 dBm | 170 mW |
| [12] | LC | 1000 MHz | 19% | 2.4% | 20 dB | 3.3 dB | 29 dBm | 34 dBm | N/A |
| [11] | LC | 1000 MHz | 10% | 2.3% | 20 dB | 0.8 dB [f] | 29 dBm | 32 dBm | N/A |
| [18] | MEMS | 155 MHz | 0.6% | 5.8% | 20 dB | 6.6 dB | N/A | 30 dBm | N/A |
| [19] | MEMS | 2500 MHz | 0.1% | 0.02% | 20 dB | 11 dB | N/A | N/A | N/A |
| [14] | MEMS | 146 MHz | 0.1% | 0.2% | 15 dB | 8 dB | -8 dBm | N/A | N/A |
| [27] | MEMS | 1165 MHz | 0.1% | 0.3% | 15 dB | 12 dB | N/A | N/A | N/A |
| **This work** | **SAW** | **900 MHz** | **1.4%** | **2.6%** | **20 dB** | **2.9 dB** | **29.5 dBm** | **41 dBm** | **0.2 mW** |

[a-b] Defined by the ratio with center frequency.
[b] Defined by the IX value.
[c] Results are broadband measured from DC to 3 GHz.
[d-e] Assuming center frequency is 1.5 GHz.
[f] Baluns are de-embedd.

The output spectrum of the circuit is also measured using a spectrum analyzer with an input tone of 892 MHz and an amplitude of –20 dBm, as shown in Fig. 13. As expected, all the odd-order IMPs are suppressed, due to the use of differential configuration. Therefore, the output spectrum shows only even-order IMPs.

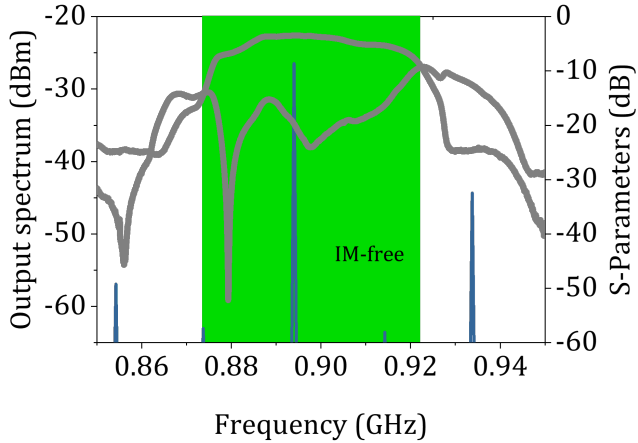

Fig. 18 Output spectrum in the non-reciprocity region. Thanks to the quad configuration, the non-reciprocal band is IM-free.

As expected, the even-order IMPs are not suppressed by the differential configuration. In order to further suppress these products, the quad configuration with schematic shown in Fig. 6 was experimentally implemented by combining two differential circuits, as shown in Fig. 14. SMA tees were used to combine the pairs of ports on two PCBs. The S-parameters of the quad configuration were then measured, using the same testing set-up shown in Fig. 10, with a terminal impedance of 25 ohm. The measurement shows a similar result of S-parameters, depicted in Fig. 15 and Fig. 16, as the differential configuration. The measured IL is 3.3 dB, which is slightly higher than differential case. This is due to the losses coming from the SMA tees and the additional cables to connect the two PCBs. The measured RL is more than 10 dB for both the Tx and the ANT ports. It is worth mentioning that even though the quad configuration is matched to 25 ohm, a 50 ohm matched quad architecture can be designed by customizing the characteristic impedance of the SAW filters.

The output spectrum of the quad configuration is measured by sending an input signal of 894 MHz with an amplitude of –20 dBm, as shown in Fig. 17. As expected, the output spectrum shows large suppression of the all the first three IMPs, i.e., the suppression is more than 37 dBc. The finite IMP suppression can be explained by the non-perfect modulation phase provided by the function generator and the slight mismatch between the two PCBs. Since the 4th-order IMPs are far away from the fundamental harmonic, an IMP-free non-reciprocal band is demonstrated in Fig. 18.

The linearity was tested by measuring the 1-dB input compression point (P1dB) and the input-referred third-order intercept point (IIP3). Fig. 19 and Fig. 20 show that a P1dB of 29.5 dBm and an IIP3 of 41 dBm were achieved. This high linearity is due to the use of large RF switches with high power handling, enabled by the low modulation frequency associated with the employed SAW filter technology.

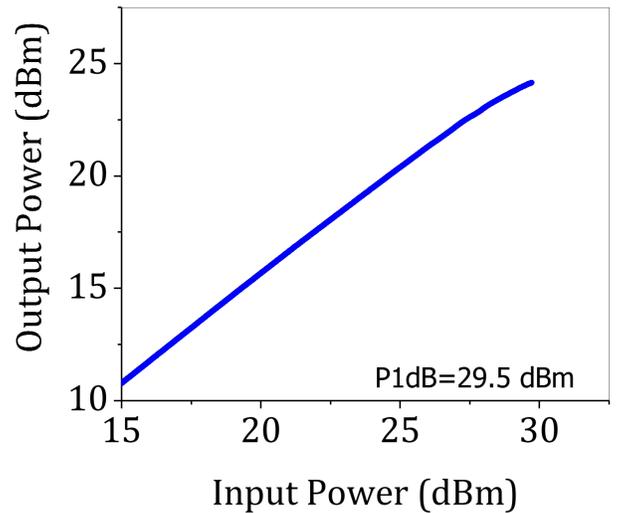

Fig. 19 Measured P1dB of the differential configuration.

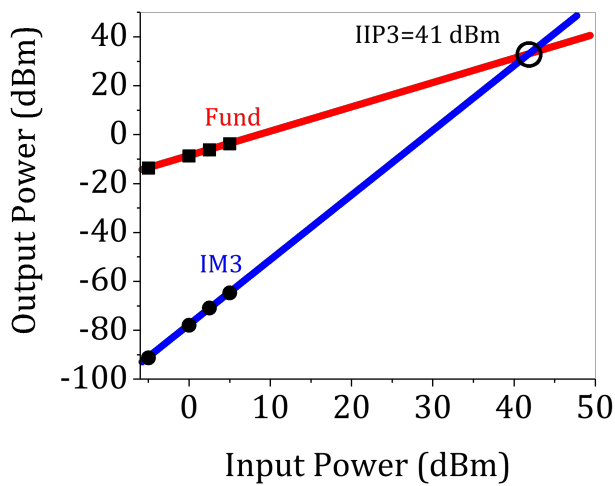

Fig. 20 Measured IIP3 of the differential configuration.

## IV. Conclusion

In this paper, we demonstrated a magnet-free circulator based on the spatiotemporal modulation of SAW filters. Thanks to the high Q factor of SAW filters, the circulator shows low IL, large IX, broad BW and ultra-low modulation frequency all at the same time. Table I summarizes the presented results in comparison to other works on magnet-free circulators. Working at a center frequency of 900 MHz with one of the lowest IL of only 2.9 dB, this work achieves a broad BW of 2.6% with a low modulation frequency of only 1.4%. Compared to [11], [12], [15], [25] and [26] based on the spatiotemporal modulation of either TLs or LCs, the modulation frequency and power consumption are significantly reduced. For example, compared to [25] based on LC delay networks with similar center frequency, modulation frequency is more than 20 times smaller, which translates to a reduction of power consumption by a factor of 850. Therefore, compared to demonstrations based on TLs/LCs, this work overcomes the problems of large modulation frequency and large power consumption, while does not show trade-offs in other performances such as IL and BW like [14], [18], [19] and [27]. Furthermore, thanks to the low modulation frequency, RF switches with high power handling are used, therefore, the demonstrated circulator shows one of the highest linearity (P1dB of 29.5 dBm and IIP3 of 41 dBm). The demonstrated high performance magnet-free circulator is an important progress towards full-duplex communication systems in the near future.

## Acknowledgment

The authors thank Dr. Charles F. Campbell from Qorvo for providing RF switches engineering samples. The authors wish to thank Dr. Roy Olsson and Dr. Timothy Hancock for the support.